\documentstyle[12pt,a4,epsfig,epsf]{article}  
\parskip=\itemsep              
\newcommand{\beq}{\begin{equation}}   
\newcommand{\eeq}{\end{equation}}   
\newcommand{\beqar}[1]{\begin{eqnarray}\label{#1}}   
\newcommand{\eeqar}{\end{eqnarray}}   
\def\eq#1{{Eq.~(\ref{#1})}}   
   
\def\npb#1#2#3{    {\it Nucl. Phys. }{\bf B#1} (#2) #3}   
\def\plb#1#2#3{    {\it Phys. Lett. }{\bf B#1} (#2) #3}   
\def\prd#1#2#3{    {\it Phys. Rev. }{\bf D#1} (#2) #3}

\def\zpc#1#2#3{    {\it Z. Phys. }{\bf C#1} (19#2) #3}

\relax   
\begin{document}   
\title{   
QCD Saturation and Photoproduction on Proton and Nuclei 
Targets\\\  
}  
  
\author{   
{ ~J.~Bartels,  
\thanks{e-mail:jochen.bartels@desy.de}~~$\mathbf{{}^{a)}}$ 
  ~E.~Gotsman, \thanks{e-mail:   
gotsman@post.tau.ac.il}~~$\mathbf{{}^{b),\,d)}}$ E.~Levin,\thanks{e-mail:   
leving@post.tau.ac.il;\,\,\,levin@mail.desy.de}~~$\mathbf{{}^{b),\,c)}}$   
  M.~Lublinsky,\thanks{e-mail:   
lublinm@mail.desy.de;\,\,\,mal@tx.technion.ac.il}~~$\mathbf{{}^{c)}}$} \\
{and\,U.~Maor \thanks{e-mail:maor@post.tau.ac.il}~~$\mathbf{{}^{b)}}$   
}\\[4.5ex]  
{\it ${}^{a)}$ II Institut f\"{u}r Theoretische Physik,}\\    
{\it Universit\"{a}t  Hamburg, D-22761 Hamburg, GERMANY}\\[2.5ex]  
{\it ${}^{b)}$ HEP Department    School of Physics and Astronomy}\\   
{\it Raymond and Beverly Sackler Faculty of Exact Science}\\   
{\it Tel Aviv University, Tel Aviv, 69978, ISRAEL}\\[2.5ex]   
{\it ${}^{c)}$ DESY Theory Group}\\   
{\it 22607, Hamburg, GERMANY}\\[2.5ex]   
{\it ${}^{d)}$ Department of Physics and Astronomy}\\
{\it University of California, Irvine, CA 92612, USA}\\[4.5ex]
}   
      
\maketitle   
\thispagestyle{empty}   
   
\begin{abstract}   
  We calculate inclusive
 photoproduction cross-sections for both proton and nuclei
targets, using a saturation model derived from an approximate solution to
the Balitsky-Kovchegov nonlinear evolution equation. This paper extends 
our hypothesis, that one can successfully use  the dipole picture and
saturation models to describe soft processes, instead  
of the soft Pomeron Regge parametrization.
Our fit is compatible with our previous fit to DIS data,
 and utilizes the 
same  phenomenological parameters as in our  paper devoted to soft 
hadronic interactions. Using the Glauber formalism, and no additional
parameters, we calculate the cross-sections for photoproduction on various 
nuclei, and  compare our  results with the relevant data.

 \end{abstract}   
\thispagestyle{empty}   
\begin{flushright}   
\vspace{-22.5cm}   
TAUP-1001\\   
DESY 03-050\\   
\end{flushright}   
\newpage   
\setcounter{page}{1}

\section{Introduction}   
\setcounter{equation}{0}

Over 
the past few years there has been much progress 
 in the successful application of perturbative QCD to DIS processes for
  values of $Q^2 \; \geq \; 2 \; GeV^2$. 
 For very small $Q^2$, 
 Regge theory (e.g.\cite{ZEUS})  provides a reasonable  
description of the data. Attempts have been made to bridge  
 the two $Q^2$ regions, using methods incorporating high 
parton densities and saturation. Models based upon these ideas have  
lead to excellent descriptions of
the DIS cross-section for all values of $Q^2$ and   $ x\; \le $  
0.01 \cite{GBW,BG,GLLM}.  The saturation hypothesis has been  successfully 
applied to $\gamma-\gamma$ scattering in Ref. \cite{BKL1}. In our recent 
paper \cite{BGLLM} we showed that  QCD-based saturation  models 
can lead to satisfactory  descriptions of  cross sections 
 in soft hadron-hadron interactions.

 A pioneering attempt to apply  the dipole picture and 
the  physics of high parton densities to  photoproduction ($Q^2$ = 0)
processes was made by Golec-Biernat and Wusthoff (GBW) within the framework
of their saturation model \cite{GBW}. 
Our goal is to investigate futher
 to what precision  high energy photoproduction 
can be described using the saturation hypothesis. We use a saturation model
based on  a solution of the nonlinear evolution equation \cite{BK}. We also
extend our analysis  to the case of nuclei targets.

The hypothesis of saturation 
refers to the interactions between partons from different cascades, which  
are omitted in the 
linear evolution equations (DGLAP and BFKL), 
and which become more significant with increasing energy. 
The parton saturation phenomenon  introduces  a characteristic  
momentum 
scale $Q_{s}(x)$, which is a measure of the density of the saturated 
gluons. It grows rapidly with energy, and it is proportional to 
 $\frac{1}{x^\lambda}$ \cite{GBW,GLLM,GLR,hdQCD,MV} with $\lambda 
\simeq 0.2$. 
Parton saturation effects are expected to be relevant
 at low values of 
$Q^2$ and $x$, where the parton densities are sufficiently large.

The starting point of this endevour is the successful fit \cite{GLLM}
within the framework of QCD,  to 
the $F_{2}$ structure 
function data for all values of $Q^2$ and $x \; \le$ 0.01.  
This was  achieved by using an approximate solution to the 
Balitsky-Kovchegov (BK) \cite{BK}  nonlinear evolution equation, and adding a 
correcting 
function to improve the DGLAP behaviour at large $Q^2$. Although  soft 
physics  was not explicitly included, agreement with experiment was found 
for all quantities associated with $F_{2}$, in particular for the 
logarithmic slope 
$\lambda\equiv \partial\ln F_2/\partial(\ln 1/x)$, where
  a value of $\lambda \; \approx \; 0.08$ was obtained 
 for very low $x$ and  $Q^2$ well below $1\, {\rm GeV^2}$, i.e. 
in the saturation region. This value is identical to that of the 
intercept of the  "soft" Pomeron, associated with the Donnachie-Landshoff (DL) 
model \cite{DL}. 
 
In this paper we  further persue the hypothesis
that colour dipoles are the correct degrees
of freedom
 in photoproduction  scattering processes
 at high energy,   
even when large transverse distances come into play. 
We adopt the well known expression for the DIS cross-sections 

\beq \label{1.1} 
 \sigma^{\gamma^{*}p}(x,Q^2) \; = \; \int d^2r_{\perp} dz 
P^{\gamma^*}(Q^2,r_\perp,z)\, \sigma_{dipole}(x,r_{\perp}), 
\eeq 
where  $Q^2$ is the virtual photon's four momentum 
squared. $P^{\gamma^*}$ denotes the probability of finding a  $q-{\bar q}$ 
colour dipole
with  transverse separation $r_\perp$ inside the photon.
$s=W^2$ is the energy squared in the photon-proton 
system and 
 z, (1 - z)  the momentum fraction taken by the quark (antiquark) 
respectively, 
$ x \; = \; \frac{Q^2}{(W^2 \; + \; Q^2)}$. 
The
dipole cross-section $\sigma_{dipole}$  describes the interaction of 
the 
$q {\bar q}$ pair
 with the proton target. The exact form of this cross-section has not yet
been determined.
GBW \cite{GBW}
proposed a very successful parametrization for $\sigma_{dipole}$.
In our
approach $\sigma_{dipole}$
  is obtained by solving the nonlinear evolution equation, 
which is derived in QCD.

In \cite{GLLM} the nonlinear evolution  equation 
 with certain approximations,
 was solved  numerically, the
initial conditions were determined
 by fitting to 
the low $x$ DIS data.   In this paper  the resulting 
parametrization for $\sigma_{dipole}$
is  adopted for the case of 
photoproduction.  In contrast to our our treatment  of the DIS 
case \cite{GLLM}, we
  need  to introduce additional nonperturbative
parameters 
to insure a finite result in the limit $Q^2\; \rightarrow \; 0$.  
This leads to
  a successful description of both the photoproduction 
and low $Q^2$ DIS data. 
Our approach for
 photoproduction is very similar to that which we employed
 for soft hadronic interactions in Ref. \cite{BGLLM}.

\par As a further test of our hypothesis we apply our model to the case 
of the photon scattering  on nuclei targets. This extension does not 
involve any additional parameters, and can thus be regarded as an 
independent check.  All available data on nuclear shadowing is successfully
reproduced for heavy nuclei.  An approach similar to 
the one presented here  for
the photon scattering  on nuclei can be found in Ref. \cite{Armesto}. In 
\cite{Armesto} a Glauber formula  with the  GBW model was used. In this 
paper, we rely
 on the model defined in Ref. \cite{GLLM}.
 
The content of the paper is as follows. In Section 2 we discuss the 
numerical solution   of the BK  equation \cite{BK}.
  We   briefly review the GBW saturation model
 \cite{GBW}, with which we compare  some of
our results.
 In 
Section 3 we present the changes that must be introduced 
to adapt the formalism for the calculation of photoproduction 
cross-sections. Section 4 is 
devoted to the overall picture, including comparison of the model's 
results with experimental data. We extend our approach to nuclei 
targets in Section 5. 
 Section 6 contains a discussion of our 
results and our conclusions.


\section{ Description of the Saturation Models}   
  \setcounter{equation}{0} 
 
 In \cite{GLLM} an approximate solution to 
the BK non-linear 
evolution equation \cite{BK} was obtained using numerical techniques.  
 Below we  briefly review  
the method used and the main results obtained. For more 
 details of the method of solution  we refer to \cite{GLLM}. 
 
\begin{figure}[htbp]   
\center   
\epsfig{file=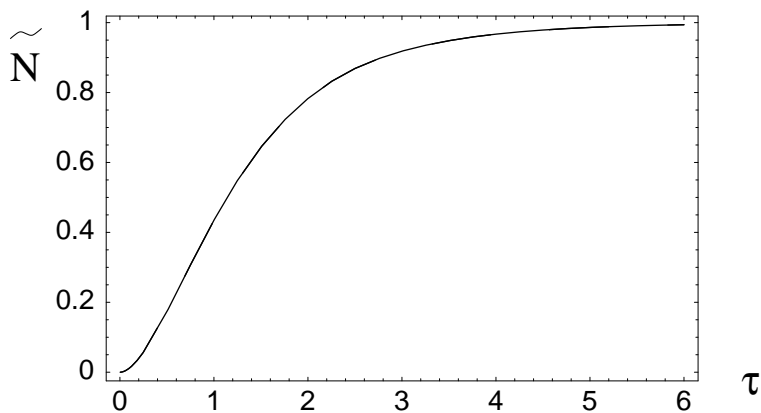,width=100mm}     
\caption{ \it $\tilde N(b_{\perp}=0)$  versus $\tau=r_\perp\, Q_s(x)$. } 
\label{scale}   
\end{figure}

The solution of the BK equation which we denote by  
$\tilde N$, takes into account the collective phenomena of high parton  
density QCD.  Starting from an initial condition which contains  
 free parameters, we have numerically solved the nonlinear evolution    
equation, restricting ourselves to the point $b_{\perp} =  0$. The 
parameters which appear in the initial conditions, have been determined by 
fitting to the $F_2$ data \cite{GLLM}, and the resulting approximate 
solution is   displayed in Fig. \ref{scale}. 
 
The $b_\perp$-dependence of the solution is restored using the ansatz: 
\beq \label{AN}  
\label{Nb} \tilde N(r_\perp,x; b_\perp)\,=\,   
(1\,-\,e^{-\kappa(x,r_\perp)\, S(b_\perp)})\,,  
\eeq  
where $\kappa$ is related to the $b_\perp=0$ solution   
\beq \label{kappa}   
\kappa(x,r_\perp)\,=\,-\,\ln(1\,-\,\tilde N(r_\perp,x,b_\perp = 0)).  
\eeq   
A Gaussian form for the profile function in impact parameter  
space was taken  as 
\beq\label{sb} 
    S(b_{\perp}) \; = \; \frac{1}{ \pi\; R_{proton}^2} 
exp(- \frac{b^2_{\perp}}{ R_{proton}^2})\,. 
\eeq 
where $R_{proton}^2\;$ =  3.1 $GeV^{-2}$ refers to the radius of the target 
proton.
Our choice of a Gaussian profile function simplifies the numerical 
calculations. It was shown  \cite{GLLMN1} that  calculations pertaining
to total and integrated cross-sections are insensitive to the details of the 
profile function, unlike the case for differential quantities.

 The dipole-proton cross-section (from eq.(\ref{DN})   
of \cite{GLLM}) is given by: 
\beq 
 \label{DN} 
\sigma_{dipole}(r_\perp,x) \; = \; 2 \; \int \; d^2b_\perp\, 
 \tilde N( r_{\perp}, b_{\perp}, x)\,.  
\eeq  
The contribution of the $F_2$ structure function is
  \beq \label{F2T} \tilde F_2(x,Q^2)\,\,\,=\,\,\frac{Q^2}{4\pi^2  
}\,\int\,\,d^2 r_{\perp} \int \,d z\,\, P^{\gamma^*}(Q^2;  
r_{\perp}, z) \,\,\sigma_{\rm dipole}(r_{\perp}, x)\,\,. \eeq  

The physical content of \eq{F2T} is easily explained. 
It describes the  two stages of  
 DIS \cite{GRIB}. The first stage is  
the decay of a virtual photon into a colorless dipole ($ q \bar q  
$ -pair), the probability of this decay
 is given by  
$P^{\gamma^*}$.  
 The second stage is the interaction of the dipole  
with the target i.e. ($\sigma_{\rm dipole}$ in \eq{F2T}). This equation  
is a simple manifestation of the fact that color dipoles are the  
correct degrees of freedom in QCD at high energies \cite{MU94}.  
 The probability $P^{\gamma^*}$ is given by the 
square of the QED wave functions  
of the virtual photon, which are well known  
\cite{MU94,DOF3,WF}:  
  
\begin{eqnarray}\label{psi}  
& & P^{\gamma^*}(Q^2; r_{\perp}, z) \,=\,\frac{N_c}{2\,\pi^2}  
 \sum_f  \,Z_f^2\,\,\times \\
& &\left\{(z^2+(1-z)^2)\,a^2\,K_1^2(a\,r_\perp)\,\,+\,\, 
m_q^2\, K_0^2(a\,r_\perp)\,\,+ \,\,
 4\,Q^2\,z^2\,(1-z)^2\,K_0^2(a\,r_\perp)\right\}\, ,  \nonumber
 \end{eqnarray} 
with $a^2=Q^2 z (1-z)\,+\,m_q^2$.  
  
In Ref. \cite{GLLM} we assumed the mass of the quark to be zero i.e.
we took  $m_q=0$ in the above equation.
 The structure function $F_2$ is given by a sum of  
three contributions:  
  
\beq\label{F2} F_2\,=\, \tilde F_2\,+\,\Delta F\,+\,F_2^{NSQ}\,,  
 \eeq  
where the first term is given by Eq. (\ref{F2T}).
The second term is a DGLAP (large $Q^2$) correction to $\tilde F$ and is
of no importance for the present analysis. These two 
terms     take into account  the gluon  
contribution to $F_2$.  Gluons are related to the singlet  
quark distributions. The third term in (\ref{F2}) includes  
 contributions of non-singlet quark distributions:  
  
\beq\label{NSQ} F_2^{NSQ}\,\,=\,\,\sum_{i=u,d}\,\, e_i^2\,  
\,q_i^V\,. \eeq  
  
At the present stage of our investigation we "borrow" the valence quark  
distributions ($q_i^V$) from the LO CTEQ6 parametrization \cite{CTEQ}. It is  
important to note  that these distributions decrease with  
decreasing $x$, and are  practically of no significance below  
$x\simeq 10^{-3}$.
As the valence quark parametrization is not given below some scale of 
order
$1\,GeV^2$, for low photon virtualities  we freeze the valence quark 
distribution  at  its $Q^2=1\,GeV^2$ value.

\par A  popular saturation model was proposed by 
GBW \cite{GBW}. In this 
model the dipole cross-section is taken as
\beq\label{GW2} \hat{\sigma}_{dipole}(r_\perp,x) \; = \;
\sigma_{0}\left[1 \; - \; exp(- \frac{r^2_{\perp}}{4R^2_{0}})\right]
\eeq
 , with $R^2_0(x)[GeV^{-2}] \; = \; (\frac{x}{x_0})^\lambda$.
 The values of the parameters which were determined by fitting to DIS data
at HERA for $x \; \leq \; 0.01 $, are: $\sigma_0 \; = \; $23 mb,
$\lambda \; = \;$ 0.29 and $x_0 \; = \; 3 \times 10^{-4}$. 
The $r_\perp$ dependence was taken as Gaussian, which leads to a 
constant-cross section $\sigma_{0}$ for large $r_\perp^2$ (or small 
$Q^2$). 

\par The dipole cross-section of the GBW saturation model and 
$\tilde N$ are closely related. Both models take into account gluon
saturation, preserve unitarity, and describe the physics associated with 
"long distances". Whereas $\tilde N$ has its origins in QCD, 
$\hat{\sigma}_{dipole}$  is phenomenological.

\par There are  differences between the two models. Unlike the GBW
$ \hat{\sigma}_{dipole}$, the dipole cross-section obtained from the 
solution of 
the BK equation is not saturated as a function of $x$. 
 With the assumed Gaussian profile
function, the integration over impact parameter $b_{\perp}$
  leads to a logarithmic growth with decreasing $x$.


\section{Details of  Our Calculation} 
\setcounter{equation}{0} 
 
In Ref. \cite{GLLM} we considered a massless ($m_q=0$) case only. 
 To proceed to the limit $Q^2\rightarrow 0$ we need to introduce a 
finite
mass as a cutoff for the $r_\perp$ integration in Eq. (\ref{F2T}). 
 It is important to stress, however, that we do  not modify
the dipole cross section $\sigma_{dipole}$  (\ref{DN}), which was 
constructed in \cite{GLLM} from a fit to the $F_2$ data.
Since 
$x$ is not defined 
 for  photoproduction,  to   relate $x$ to the energy of 
the process we  need to introduce an additional 
nonperturbative scale, denoted as $Q_0^2$. Though these two parameters are
not obviously related, we set $Q_0^2=4\,m_q^2$. We redefine $x$ as
\beq\label{newx}
x=(Q^2+Q_0^2)/W^2\,;\,\,\,\,\,\,\,\,\,\,\,\,\,\,\,\,\,\,\,\,\,\,s=W^2\, ,
\eeq
and use $m_q$ as a fitting parameter for  high energy photoproduction.
The energy  
dependence of the photoproduction cross section  
enters only through the $x$-dependence of the dipole cross section, 
the later being adjusted or constructed to describe DIS data of the $F_2$  
structure function.

Note that having introduced these modifications we alter  the results
in low $Q^2$ domain. This can potentially spoil our original fit \cite{GLLM}.
Thus  photoproduction should be considered  
 simultaneously with DIS. The fit to the
later should not  be affected by the introduction of new parameters. 
 
To obtain a finite result in the limit $ Q^2 \; \rightarrow \; 0$,
GBW \cite{GBW} introduce the quark mass as a regulator, and
 assume a common mass of 140 MeV for the three light quarks,
"which  leads to a reasonable prediction in the photoproduction region".
We refer the reader to \cite{GBW} for details of the GBW model, in this
paper we  use the GBW model only to compare with our results for 
photoproduction on  nuclear targets. 

In the colour dipole picture, 
one can hopefully  reproduce only the asymptotic energy dependence  i.e. 
the  
Pomeron contribution. The solution to the BK equation is also
 used to model this
vacuum exchange channel  in the nonperturbative domain. 
 Thus  parton-hadron duality is used in this approach.
At lower energies where most of the photoproduction 
data are available, there are also  contributions  
from nonvacuum exchanges. The evaluation of these contributions is 
beyond  the scope of our model. 

In \cite{GLLM} the contribution of the nonsinglet exchange
was represented  by the valence quarks. Since the latter
were frozen at $Q^2=1\, GeV^2$, their  contribution to the cross section 
would have
the  wrong  dependence at $Q^2\rightarrow 0$ (become infinite). An  
improved form  for this
contribution of the valence quarks below $1\,GeV^2$ is:
\beq\label{quark}
 F_2^{NSQ}\,\,=\,\,\frac{(Q^2)^{1+\beta}}{1\,(GeV^2)^{1+\beta}}\,
\frac{(1\,GeV^2\,+\,\mu^2)}{Q^2\,+\,\mu^2}
\sum_{i=u,d}\,\, e_i^2\,  \,q_i^V(Q^2=1\,GeV^2)\, .
\eeq
 Vector current conservation requires $\beta\ge 0$,
in principle,  both $\beta$ and $\mu$
are fitting parameters. This freedom was not used in 
\cite{GLLM} where $\beta=0$ and $\mu^2\rightarrow 0$. The energy dependence
of the valence quark contribution enters through its $x$-dependence, 
 which is frozen at $Q^2=1\,(GeV^2)$.
 Taking
 $\mu$ as a fitting parameter produces a good description of 
both the DIS and 
photoproduction data (keeping $\beta=0$). 
 
In our approach we use parton-hadron duality in two ways.
 For the  vacuum exchange we assume that the  
 saturation result, which was derived for large $Q^2$ virtualities, 
  is a valid  description (on average) of the cross section at small 
values
of $Q^2$. For the contribution of the non-singlet structure function  
we can use a parametrization that is successful
 for soft interactions, namely the 
exchange of  secondary Reggeons. We believe that this exchange can be 
used for larger $Q^2$ (up to $1\,GeV^2$). In 
the spirit of Ref. 
\cite{DL,GLMN}
and using Gribov`s formula \cite{GRIB} the secondary Reggeon contribution is
\beq\label{Regge}
\left(\frac{s}{s_0}\right)^{\alpha_R} \,\int\frac{dM^2\,\rho(M^2)}{Q^2\,+\,M^2}
\,g_R(M^2)\,,
\eeq 
where $\rho(M^2)$ is related to the cross section for $e^+ e^-$ 
annihilation while
$g_R$ is a residue of the secondary trajectory. We approximate (\ref{Regge})
by the following contribution
\beq\label{st}
\sigma^{\gamma\,p}\,\sim\,f(0)\,\frac{\tilde 
M^2}{(Q^2+ \tilde M^2)}\,(s/s_0)^{\alpha_R}\,;
\,\,\,\,\,\,\,\,\,\,\,\,\,\alpha_R\,=\,-0.45\,;\,\,\,\,\,\,\,\,\,\,\,\,\,
s_0\,=\,1\,(GeV^2)\,.
\eeq
 $f(0)$ is the residue at  $Q^2=0$, and
is determined by fitting to the low energy photoproduction data.
$\tilde M$ is an additional fitted parameter.
\section{Photoproduction and DIS at low $Q^2$} 
\setcounter{equation}{0} 
 
A fit to high energy photoproduction  data yields the value of 
$$m_q\,=\,0.15\,GeV \,.$$
 Note that we obtain  practically the same quark mass as GBW \cite{GBW}.
 
\par
\begin{figure}[htbp] 
\center    
\epsfig{file=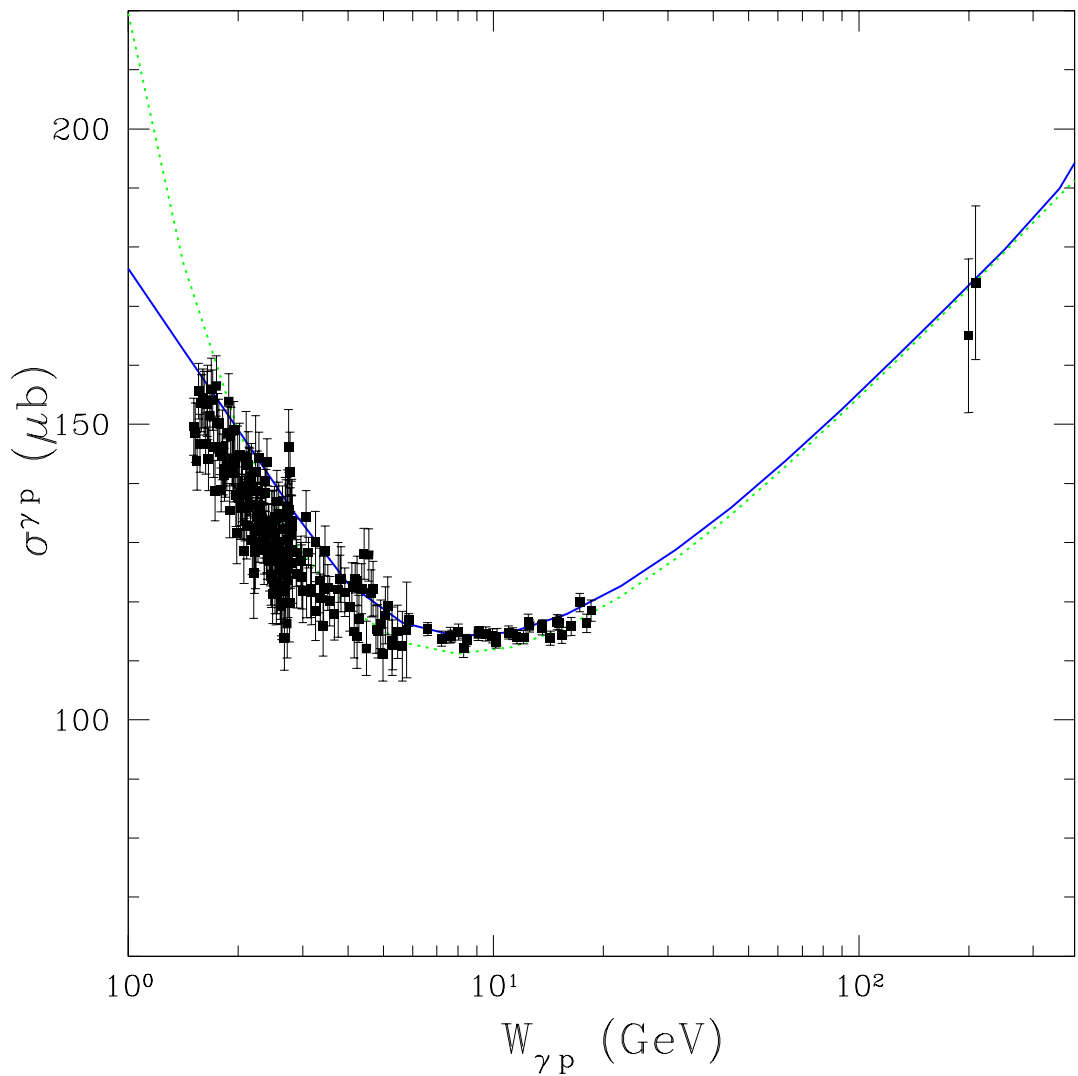,width=150mm}    
\caption{Photoproduction as a function of $\gamma-p$ energy.
Preliminary data from \cite{ZEUSphoto}. The solid line indicates the
results of Model VQ, while the dashed line those of Model ST.} \label{phot}    
\end{figure}

\begin{figure}[htbp]
\center     
\epsfig{file=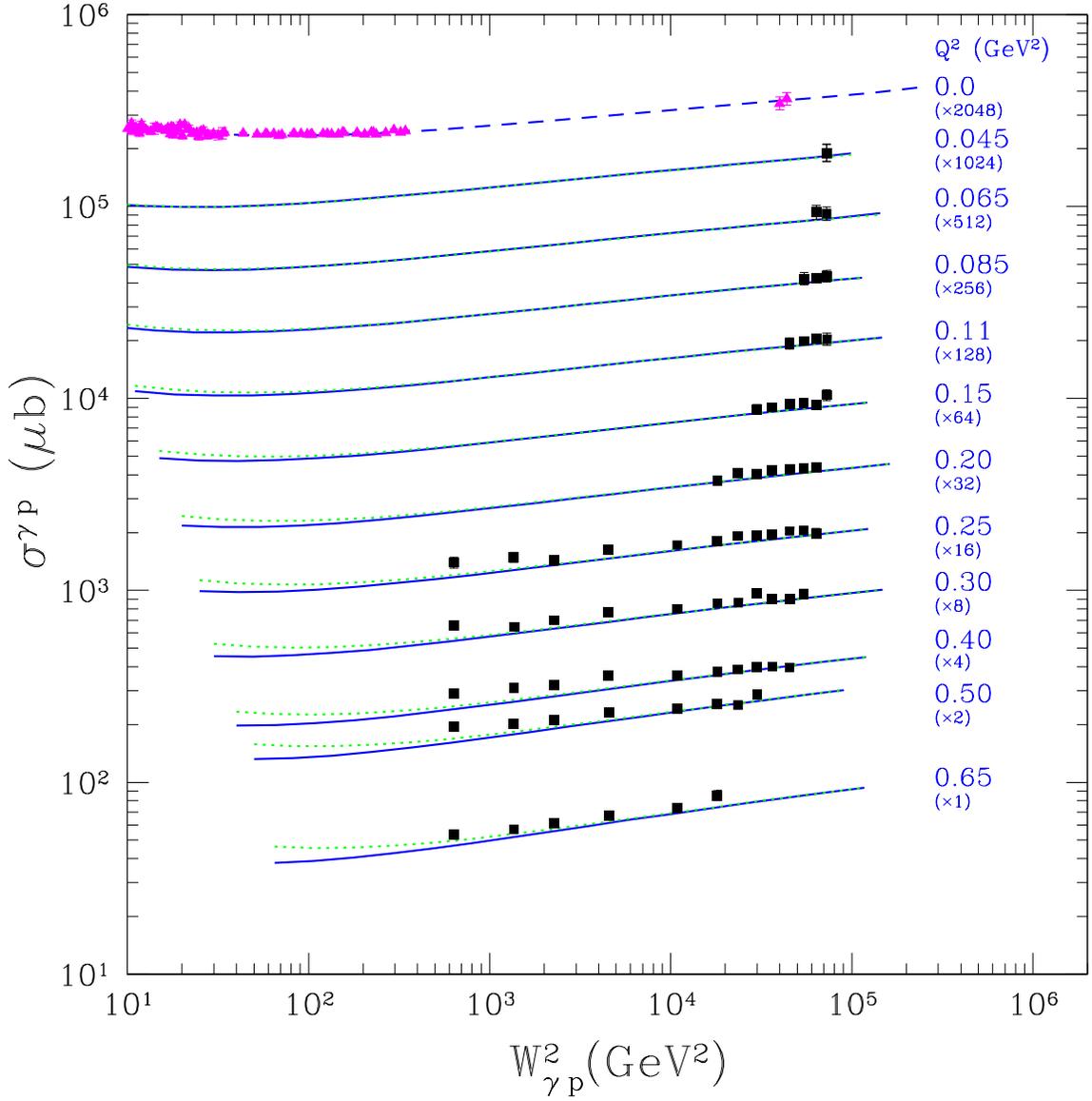,width=160mm}      
\caption{ DIS cross section at very low $Q^2$
 The two curves correspond to two different models for the secondary
trajectory (see text). Data from \cite{ZEUSlowx}.} \label{phot1}      
\end{figure}

 As most of the  photoproduction data was measured at lower 
energies, it is essential to include an additional component which
is dominant in this energy range. We attempt two alternative 
parametrizations for this component.\\ 
The first is based on the Regge pole model Eq.(\ref{st}) which we denote:

\begin{itemize}

\item {\bf Model ST:} In this model
 there are two  parameters to be fitted  which determine the 
contribution of the non-leading (or secondary trajectories) to the
DIS cross-section (see Eq. (\ref{st})). These are
$$f(0)\,=\,0.19\, mbarn \,;\,\,\,\,\,\,\,\,\,\,\,\,\,\,\,\,\,\,
 \tilde M^2\,=\,2\, GeV^2 \,.$$
The cross section obtained with these parameter values is
 indicated by the dashed curve in Figs. \ref{phot}, \ref{phot1}.
The second parametrization is based on the quark model: 

\item{\bf Model VQ:}
 This  alternate method of accounting for
the non-singlet contribution to the DIS cross-section is via the valence
quarks (see Eq. (\ref{quark})).  There are also two free parameters here, 
whose values are determined by fitting to the data.
 These are 
$$\beta=0\,\,\, (kept \,fixed) ; \,\,\,\,\,\,\,\,\,\,\,\,\,\,\,\,
\mu^2\,=\,0.13 \,GeV^2 \,.$$
The resulting photoproduction cross section with this parametrization
 for the valence quarks is shown as the full line in Figs. \ref{phot}, 
\ref{phot1}.
\end{itemize}

 We note an interesting duality between the two models, both produce 
reasonable descriptions of the experimental measurements 
(see Fig. \ref{phot}), 
 including the low energy photoproduction data.
  For the case of a proton target we do not present a comparison 
with the GBW model, since the later was formulated only for high energies 
and does not include any contribution of secondary trajectories.

\section{Photoproduction on Nuclei}
  In our approach we obtain a very small  effective radius of the
nucleon. The consequence  of such a small radius is that the 
typical parameter
that governs the dipole interaction with the nuclear target,
is also very small,
 namely
\begin{equation} \label{TYPA}
\pi\,R^2_{proton} \,S_A(0)\,\,\leq\,\,1\,\,\,\,\,\,\,\,\,\,\,\,\,\,\,
for\,\,\,\,A\,  \leq 200,
\end{equation}
 where $S_A(b_\perp) $ denotes the Wood-Saxon impact parameter distribution 
of the nucleons in a nucleus.
In the standard approach, $\pi\,R^2_{proton}\,S_A(0)$ is considered to be 
large.
Instead of one  equation \cite{BK}, 
\eq{TYPA} leads to a set of non -linear equations  
 which have to be formulated and solved.  Before doing so 
it is necessary to determine the $b_\perp$ distribution for  
dipole-nucleon 
scattering, which is still an ansatz in our approach.
At  present, the best that we can 
do  is to extend the Glauber formalism which takes into 
account all rescatterings inside the nucleus. This approach is certainly
correct at not too high energies, however to determine the region of 
applicability 
we need to solve the explicit set of equations.

The Glauber formula has a familiar form:
\begin{eqnarray}\label{Glauber}
\sigma(\gamma^*A \longrightarrow \gamma^*A ) = \int d^2 b_\perp
\int dz\, d^2 r_\perp P^{\gamma^*}(r_\perp,z,Q^2)\,2\,
\left( \,1\,-\,e^{- \frac{\sigma_{dipole}}{2} \,S_A(b_\perp)}
\right)\,.
\end{eqnarray}
 In our model $\sigma_{dipole}$ is given by Eq. (\ref{DN}). 
We use the GBW model for the dipole cross section for comparison
(see also Ref. \cite{Armesto} for a similar analysis).

\begin{figure}[htbp]
\center
\epsfig{file=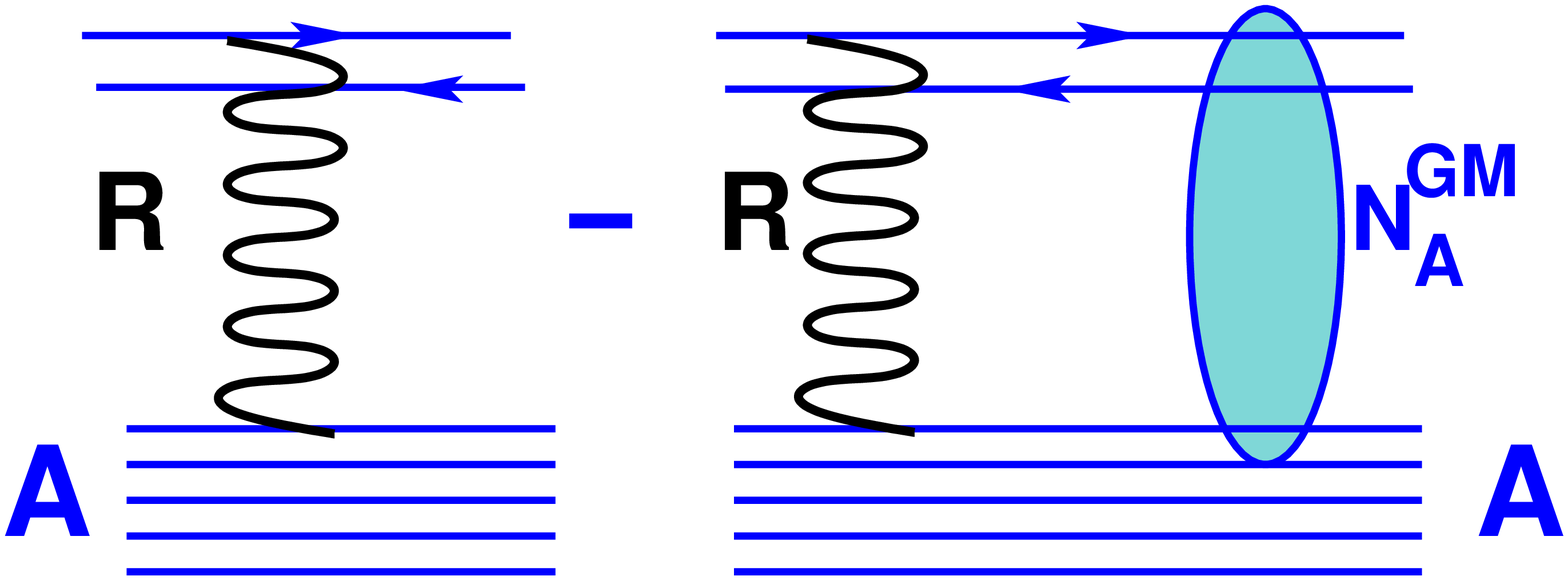,width=90mm}
\caption{The diagrams for the secondary reggeon interaction with nuclei.
$N^{GM}_A$ is the dipole-nucleus amplitude in the Glauber-Mueller 
approach. The
two diagrams cancel each other and lead to \eq{Ast}.}
\label{re}
\end{figure}

 As discussed previously, at low energies we need to add contributions
of the secondary trajectories. 
For the Glauber approach   the $A$ dependence of the secondary trajectory 
can be estimated as follows:
 
\beq\label{Ast}
A_{eff}^{st}(x) = \int d^2 r_\perp \,|\Psi_\pi(r_\perp)|^2\,
\int d^2 b_\perp\, S_A(b_\perp) \, e^{- 
\frac{\sigma_{dipole}(r_\perp,x)}{2}\,S_A(b_\perp)} .
\eeq
Here $\Psi_\pi$ denotes the pion wave function. For $\Psi_\pi$ we
use a simple Gaussian parametrization taken from Ref. \cite{Dosh}.
The physical meaning and the derivation of \eq{Ast} is clear from 
Fig. \ref{re}. Indeed, the exchange of  secondary Regge trajectories is 
screened by the black disc type interaction in the vacuum channel (the 
two 
diagrams in Fig. \ref{re} which cancel each other at small $b_{\perp}$). The 
factor $exp\left( - \sigma_{dipole}S(b_{\perp})/2 \right)$ is the 
probability that the dipole  does not to have any inelastic interaction 
(for a  more 
detailed discussion see Ref. \cite{SP}).

We compute the total cross section for $\gamma-A$ collision as a sum of
two terms 
\begin{eqnarray}\label{gA}
\sigma(\gamma^*A)\,=\, Eq.\,(\ref{Glauber})\,+\,A_{eff}^{st}\,\times\, 
Eq.\,(\ref{st}).
\end{eqnarray}

\begin{figure}[htbp]   
\center  
\epsfig{file=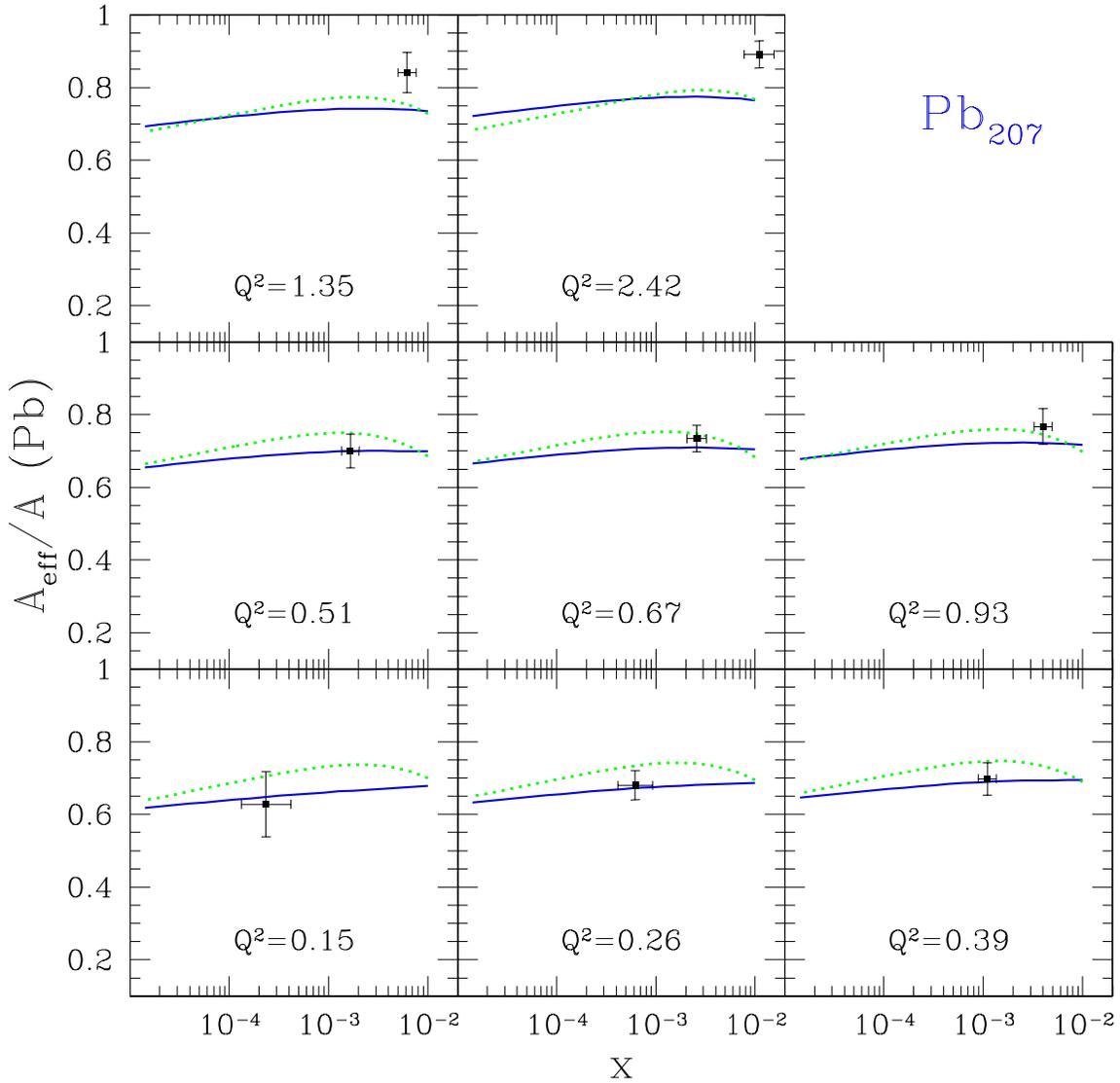,width=160mm}     
\caption{$A_{eff}/A$ on a Pb target.
The solid line denotes the results of our model including a secondary 
Regge contribution.
The dotted line is the result for the Glauber approx. for GBW model
without a secondary Regge contribution. Data from \cite{Adams}.} 
\label{aeffPb}   
\end{figure}

\begin{figure}[htbp]   
\center
\epsfig{file=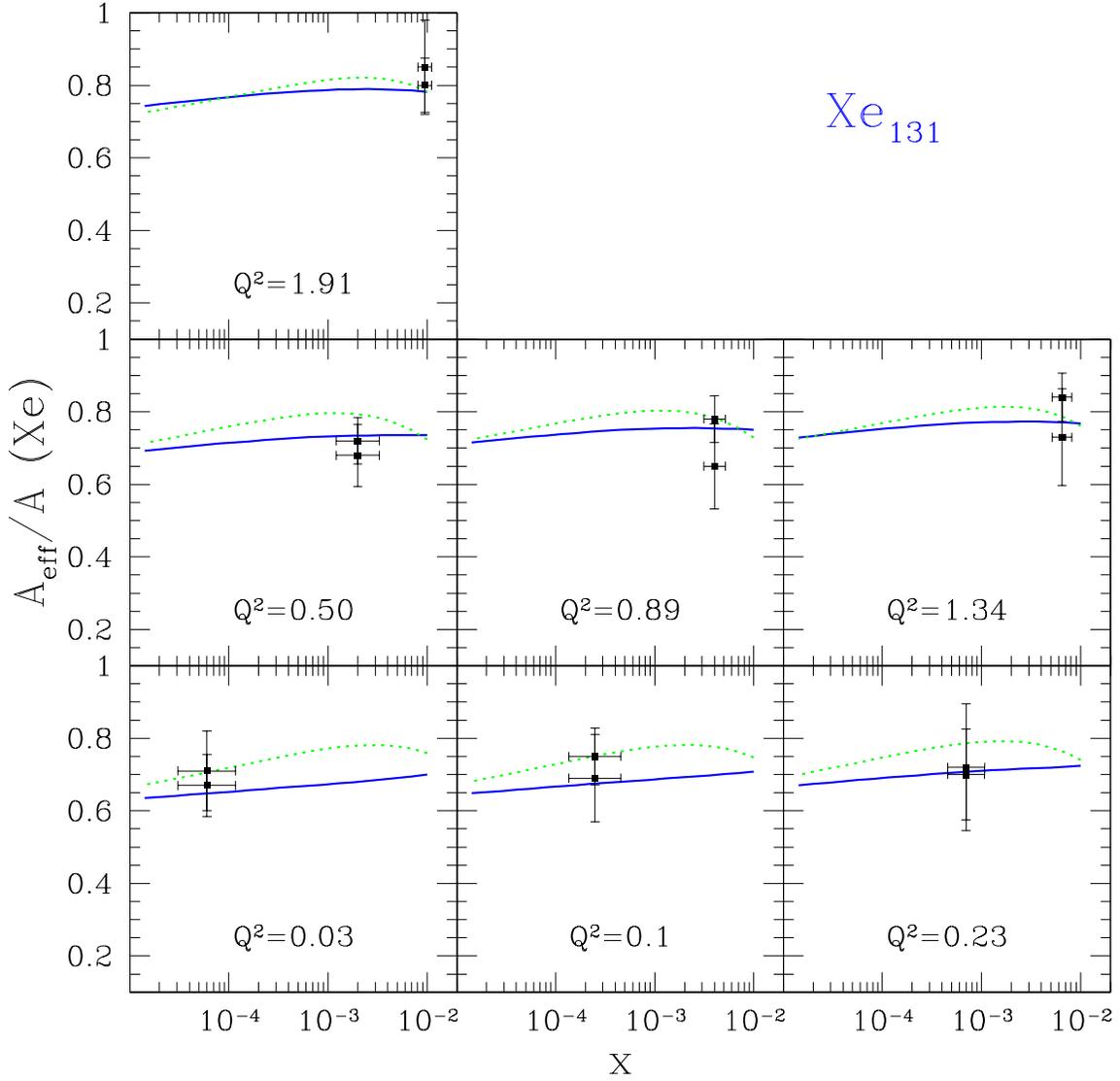,width=160mm}     
\caption{$A_{eff}/A$ on a Xe target. Notation as in Fig. \ref{aeffPb}. 
Data from 
\cite{Adams1}.} 
\label{aeffXe}     
\end{figure}

\begin{figure}[htbp]   
\center   
\epsfig{file=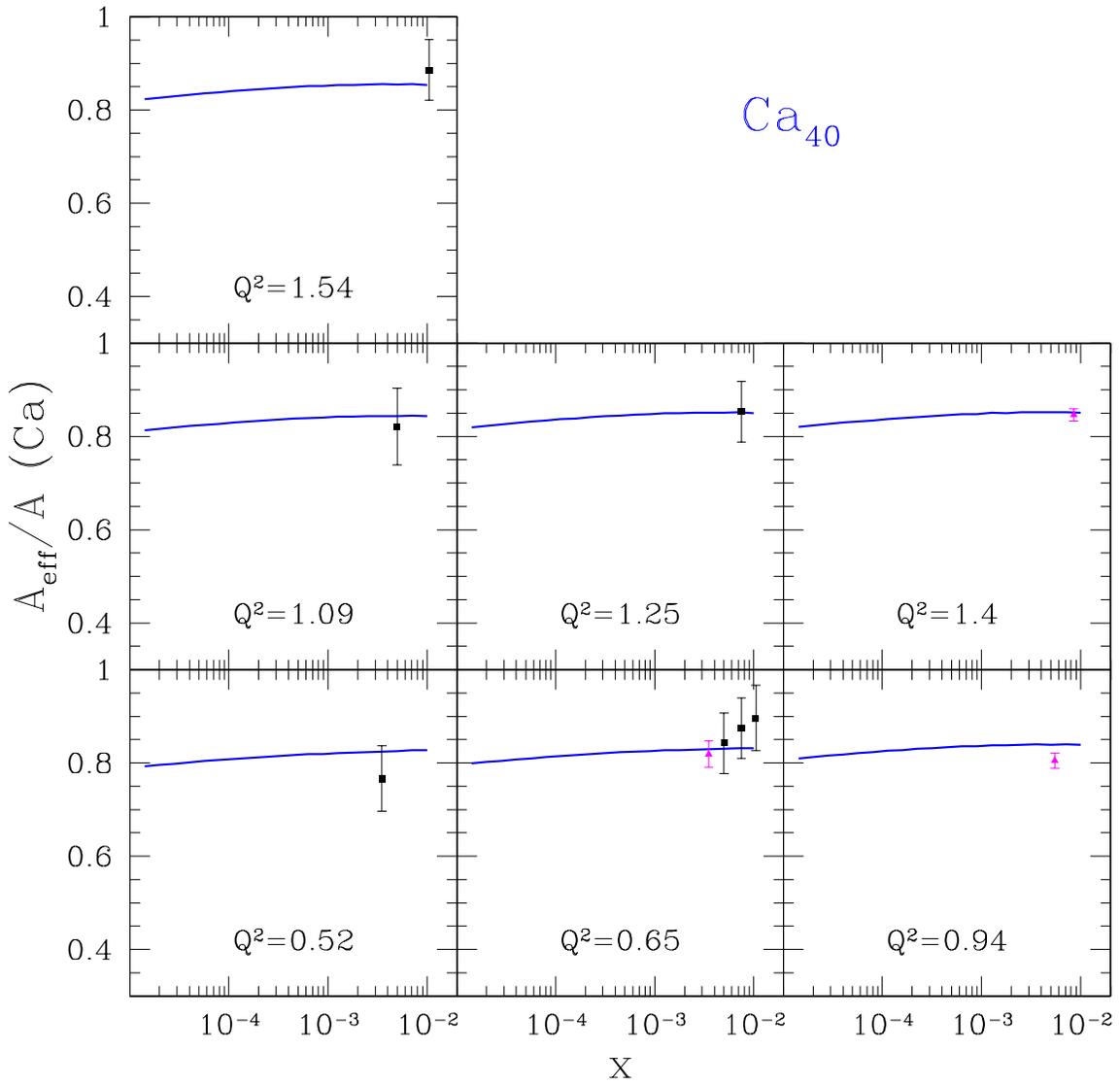,width=160mm}      
\caption{$A_{eff}/A$ on a Ca target. 
Data from \cite{Arneodo}.} 
\label{aeffCa}      
\end{figure}

\begin{figure}[htbp]   
\center 
\epsfig{file=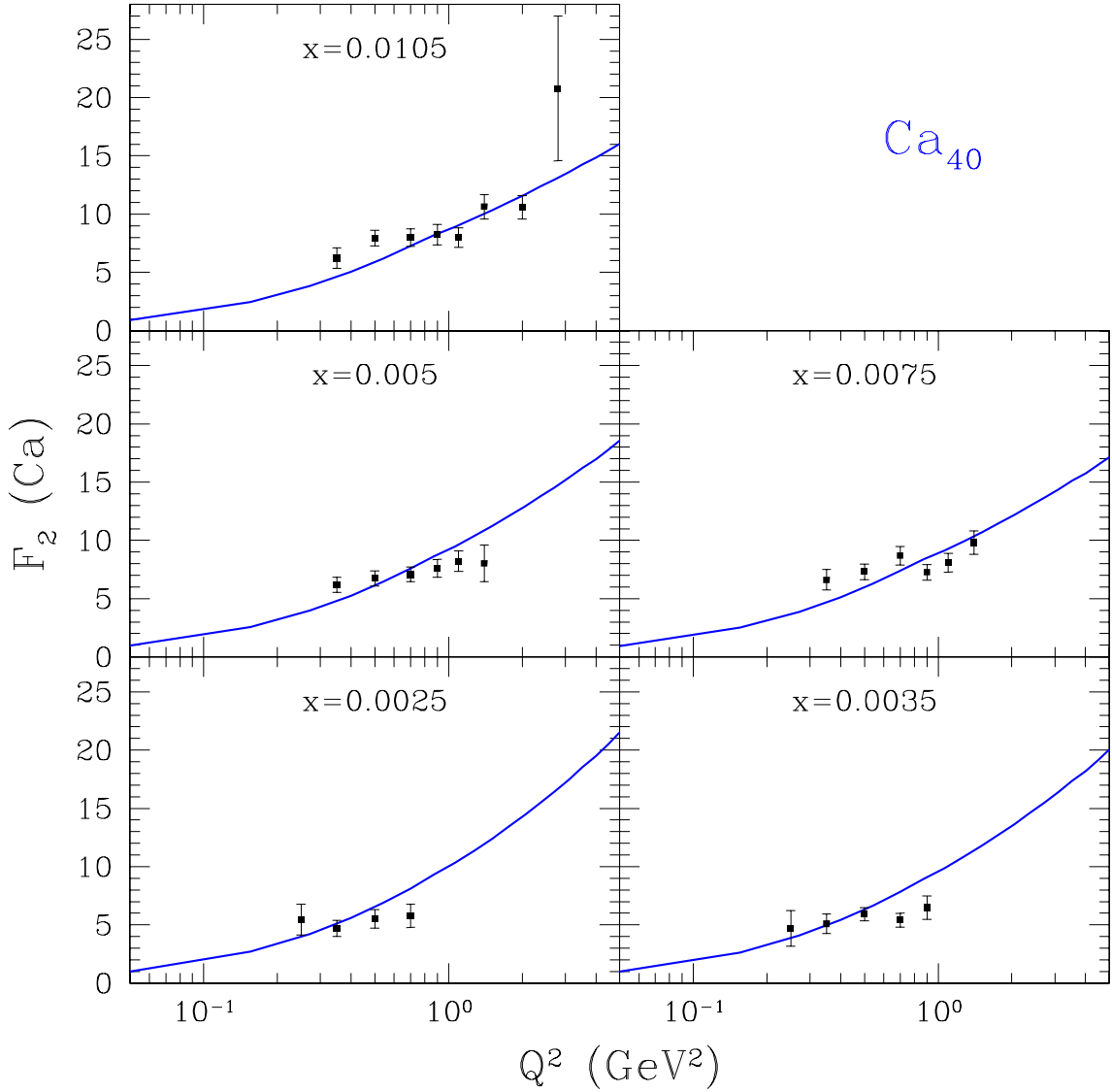,width=160mm}   
\caption{$F_2$ on a Ca target. 
Data from 
\cite{Arneodo}.} 
\label{F2Ca}   
\end{figure}

Figs. \ref{aeffPb}, \ref{aeffXe} \ref{aeffCa} illustrate the dependence of 
$A_{eff}\,\,=\,\,\sigma(\gamma^* 
A)/\sigma(\gamma^* p)$ as a function of $x$ for different values of the 
photon virtuality $Q^2$ for photo scattering on Pb, Xe, and Ca respectively.
Fig. \ref{F2Ca} shows a comparison for $F_2$ on Ca. 
 The data sets have a few percents 
overall normalization uncertainity.

 In the  figures below we use the following notation:  
the solid line shows the results of our formalism, 
where $\sigma_{dipole}$ in \eq{Glauber}
is  calculated from the solution of the 
non-linear equation.
The dotted  line illustrates the Glauber approximation for the
GBW model for the dipole-nucleon cross section, without 
a secondary Regge pole contribution. \footnote {In principle, contributions
of the secondary trajectories should be added to the GBW model as well.}  
  
Agreement with  experimental data  for both dipole models is good.
 A comparison of our model prediction for a lead target
with the results  other models 
    can be found in the recent preprint \cite{Armesto1}.

\begin{figure}[htbp] 
\center     
\epsfig{file=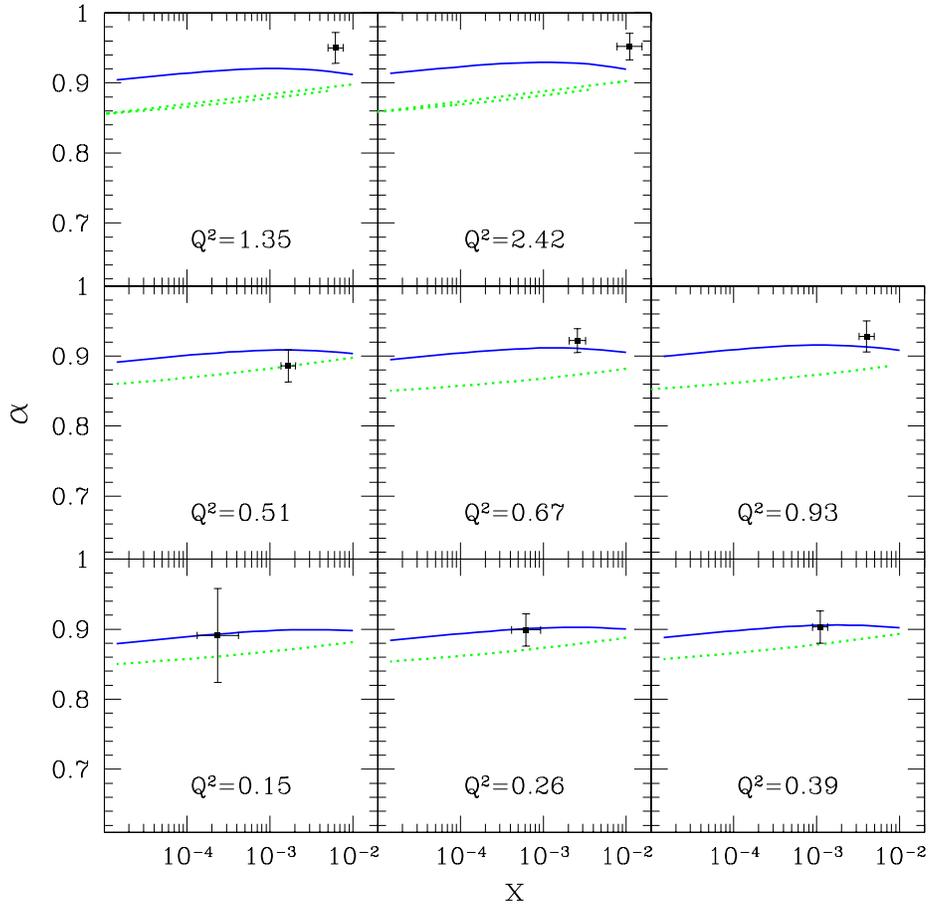,width=130mm}    
\caption{$\alpha$  as a function of $x$ for different
values of $Q^2$. Notation as in Fig. \ref{aeffPb}. Data from \cite{Adams}. }   
\label{alpha}   
\end{figure}

 In Fig. \ref{alpha}  we plot the $x$ dependence 
of $\alpha$ for the parametrization $\sigma(\gamma^* A) 
\,\propto\,A^\alpha$.  Compared to GBW 
our results appear to be in   slightly 
 better agreement with the data. However, the 
data
are not precise enough to really distinguish between the two models.

\begin{figure}[htbp]   
\center   
\epsfig{file=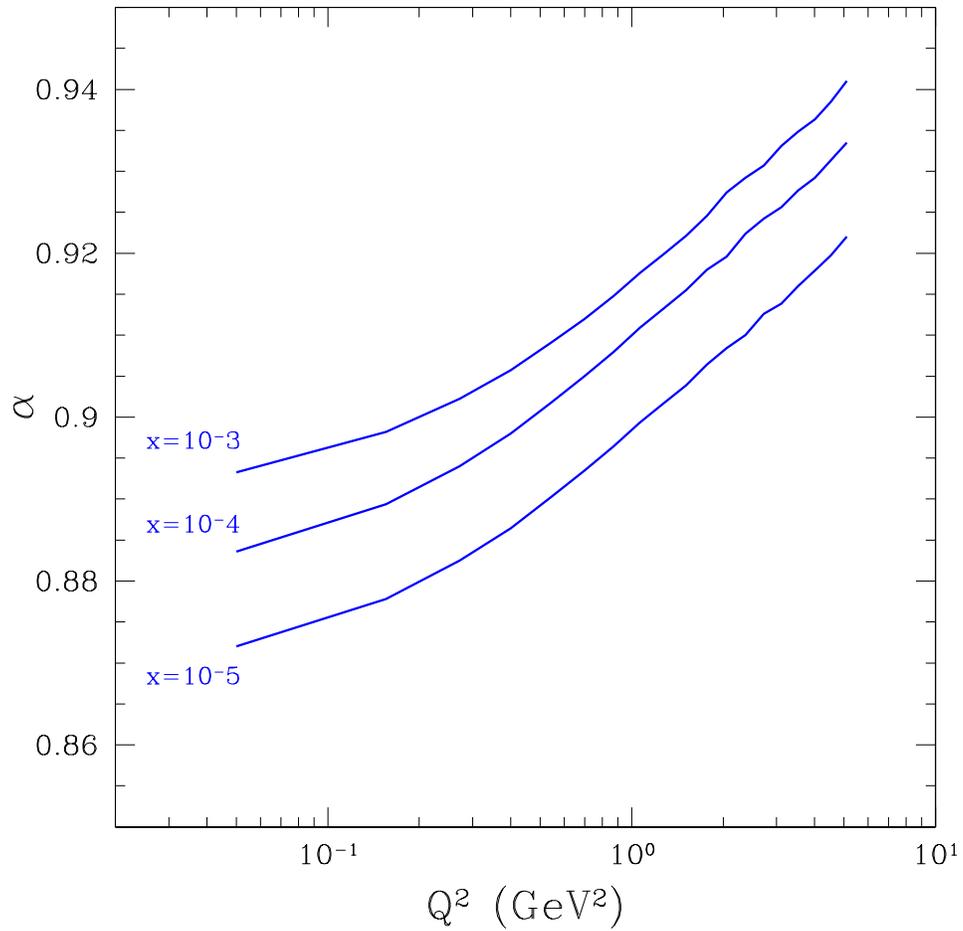,width=130mm}     
\caption{$\alpha$ as a function of $Q^2$ for fixed values of $x$.
Prediction of our model including a secondary Regge contribution.  } 
\label{alpha1}   
\end{figure}

 Our prediction for the $Q^2$ dependence of $\alpha$  is shown in 
Fig. \ref{alpha1}.

\begin{figure}[htbp]
\center   
\epsfig{file=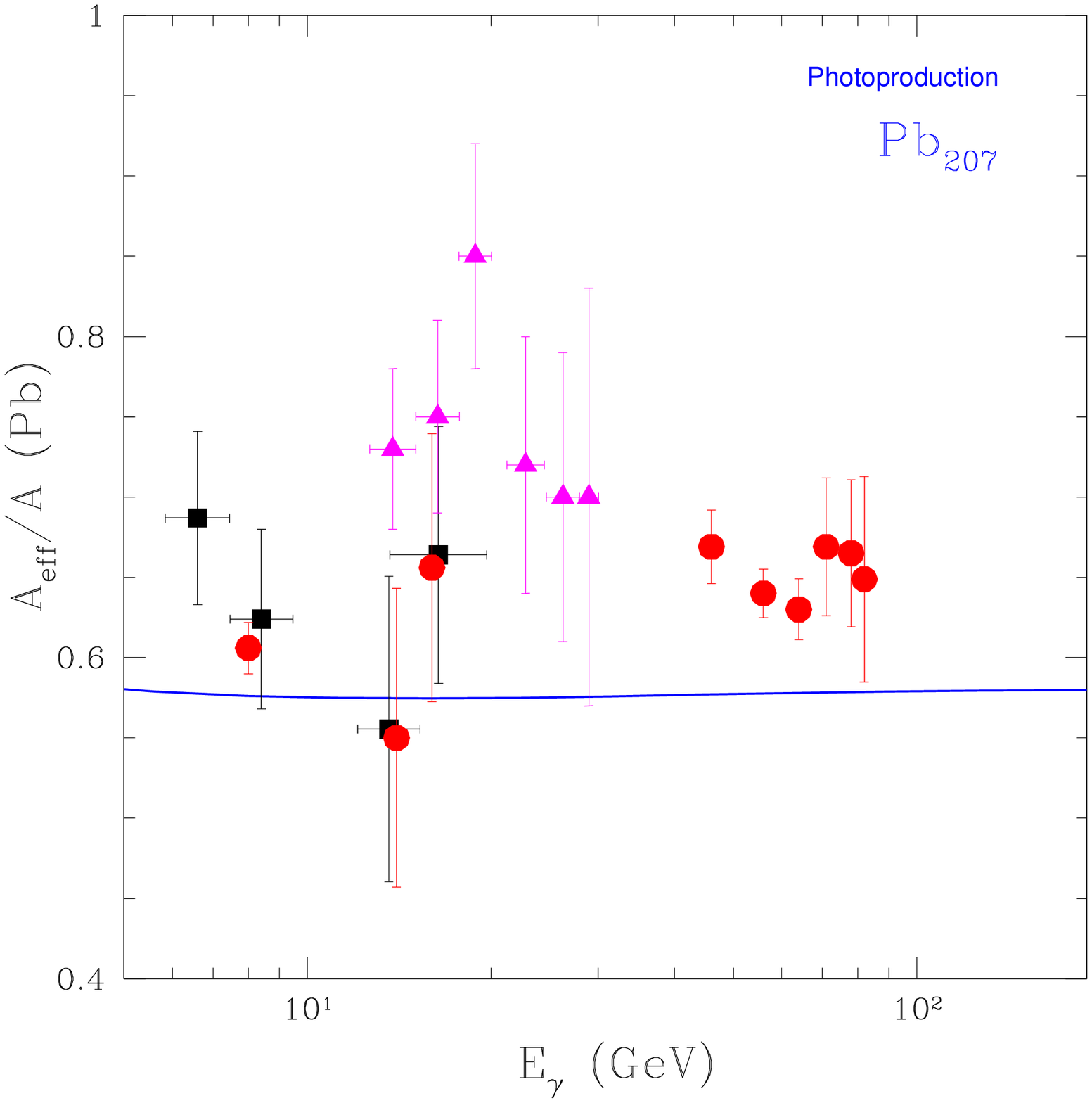,width=120mm}      
\caption{Photoproduction: $A_{eff}/A$  as a
function of $\gamma $ energy for scattering on a Pb target. 
Data from \cite{Cal1}.} \label{paeffPb}   
\end{figure}

\begin{figure}[htbp]   
\center  
\epsfig{file=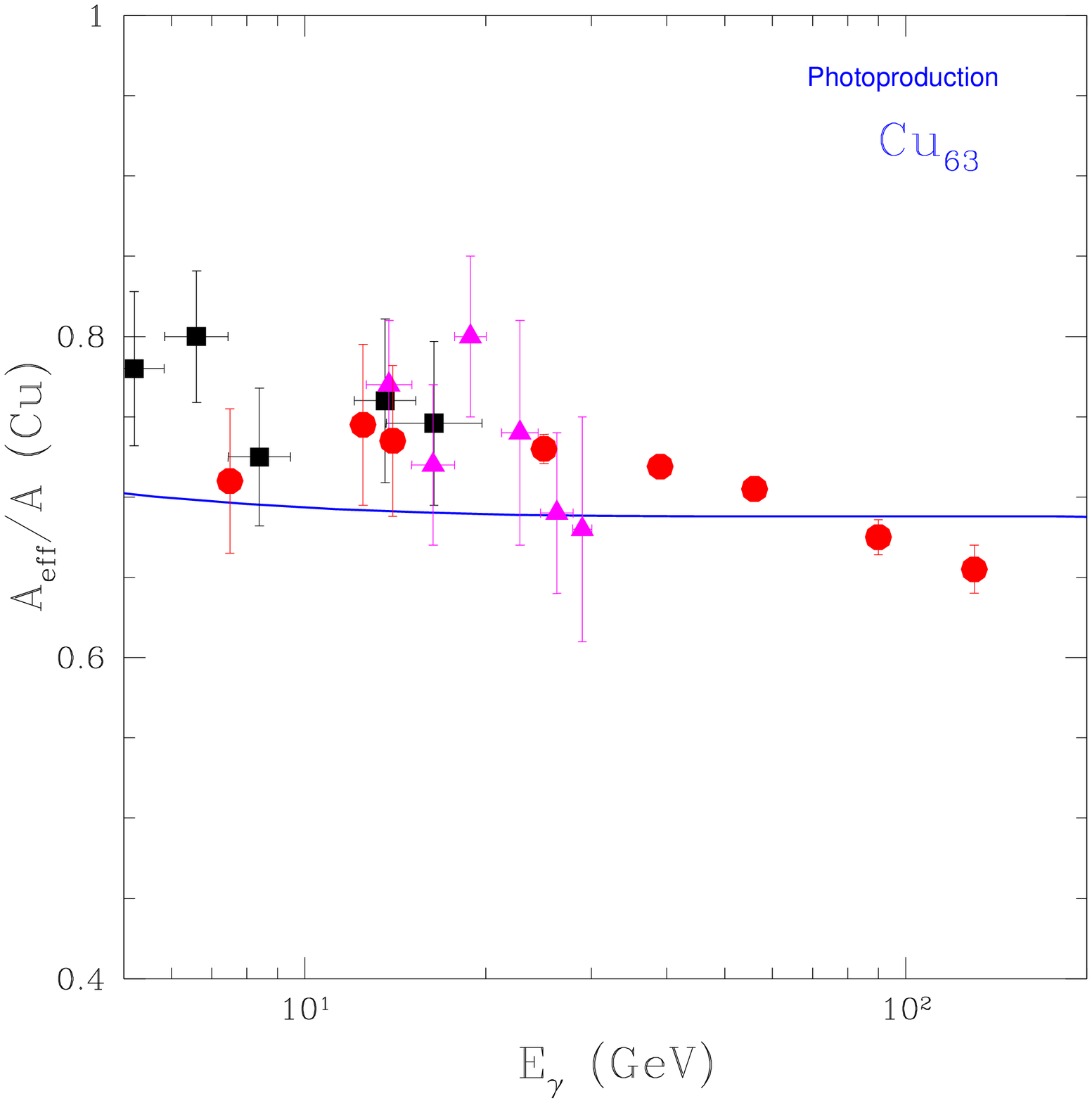,width=120mm}     
\caption{Photoproduction: $A_{eff}/A$  as a
function of $\gamma $ energy scattering on Cu target. 
Data from \cite{Cal1}.} 
\label{paeffCu}   
\end{figure}

 We compare the predictions for the energy behaviour of our approach with
the relevant data for $A_{eff}/A$ for photoproduction on Pb in Fig.
 \ref{paeffPb},
and on Cu in Fig. \ref{paeffCu}.  
We slightly underestimate the data for Pb case, and 
are consistent with the Cu measurements.
 Some points of the Caldwell data (1979) \cite{Cal1}
 are read off the plots and divided
by our photoproduction cross section.

\begin{figure}[htbp] 
\center     
\epsfig{file=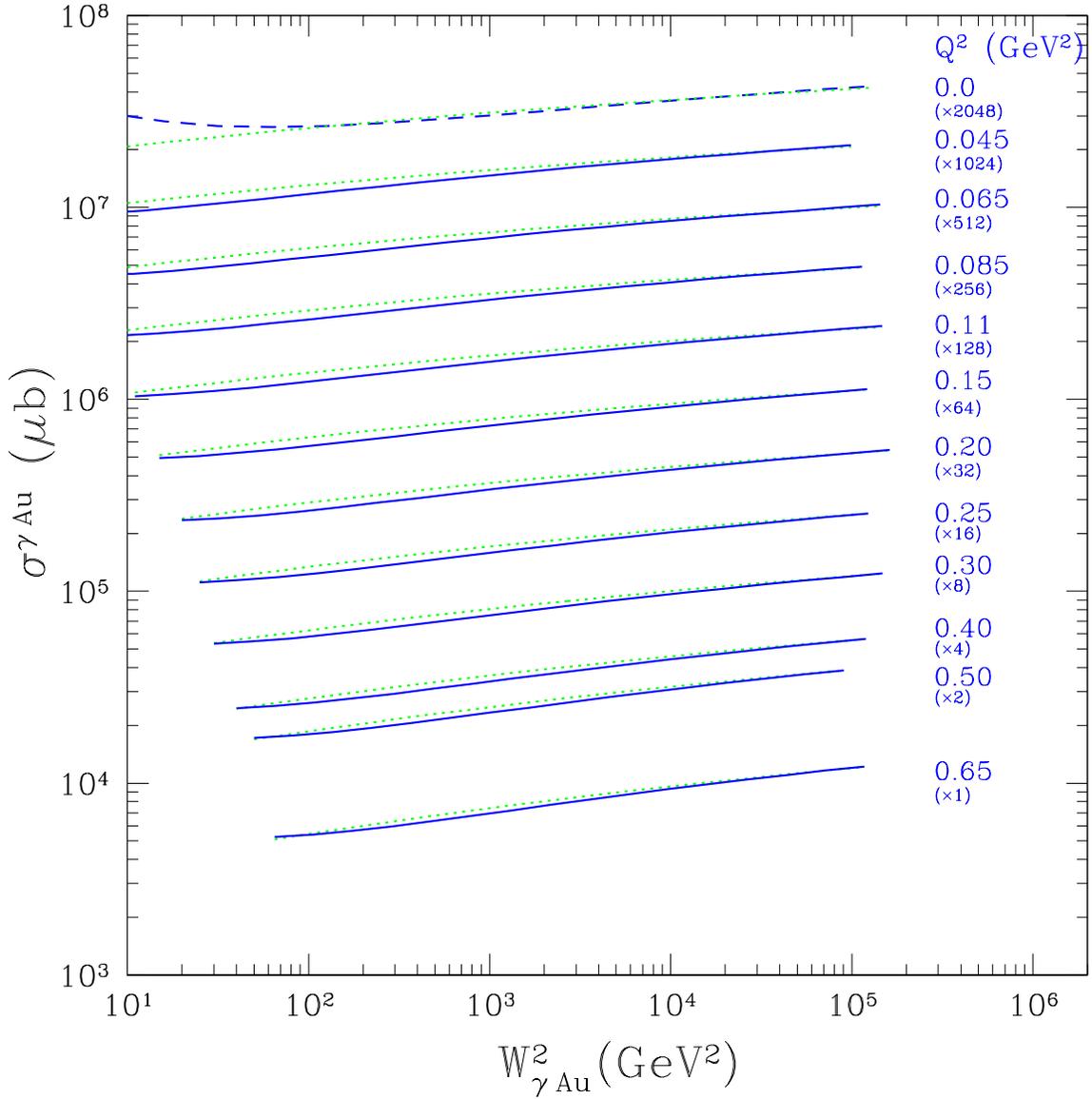,width=160mm}      
\caption{The predictions of our model for the
DIS cross section  on Au at very low $Q^2$. Notation as in Fig. 
\ref{aeffPb}.} 
\label{disAu}     
\end{figure}

\begin{figure}[htbp]   
\center  
\epsfig{file=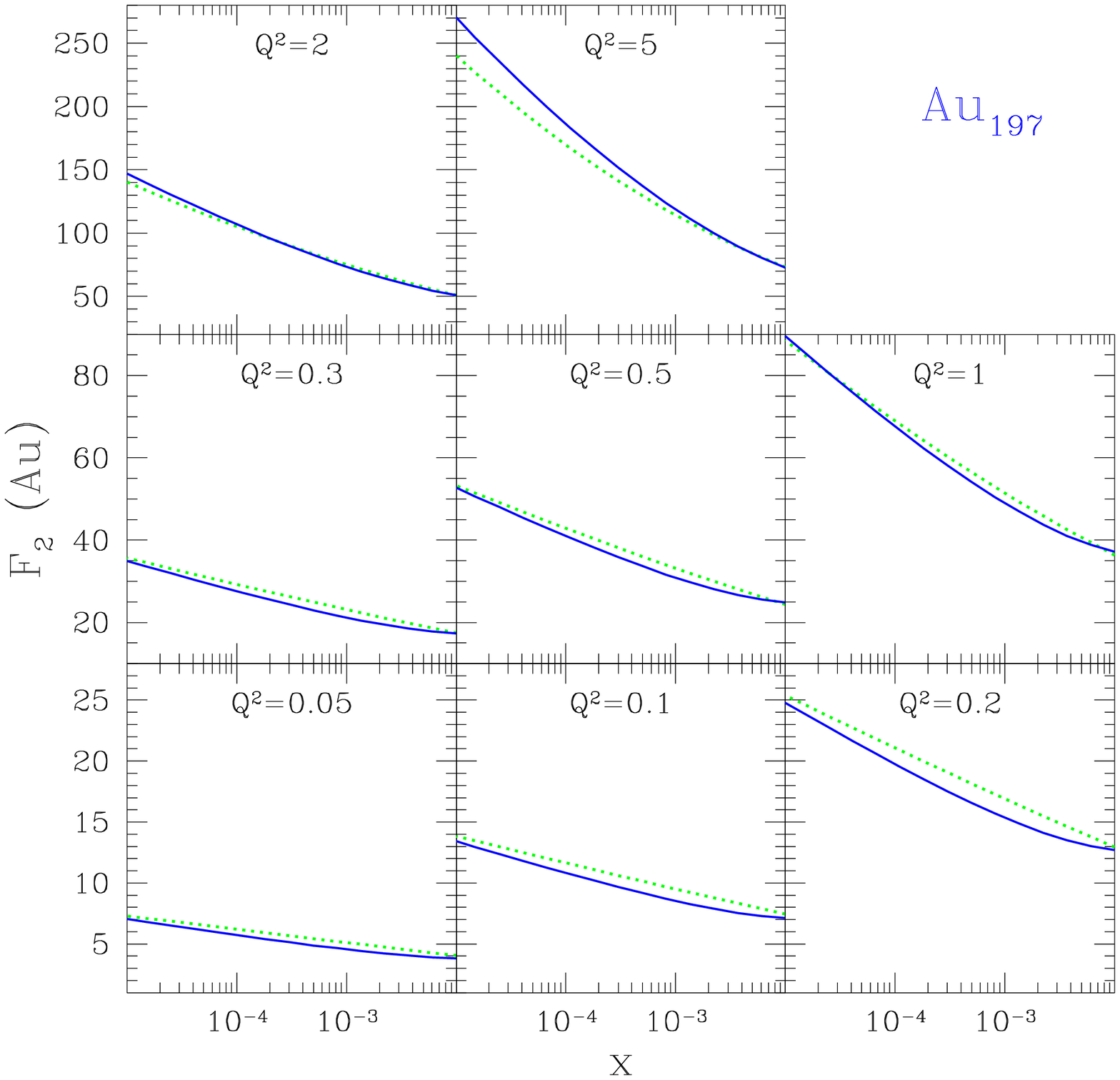,width=160mm}    
\caption{$F_2$ for gold ($A=200$) as a function of $x$ for
fixed values of $Q^2$. Notation as in Fig. \ref{aeffPb}}. \label{F2Au}   
\end{figure}

The curves in
 Fig. \ref{disAu}  which show  predictions for 
the values of the total $\gamma^* A$ cross section are crucial as a
check of  our approach.
 The results of both models are very close  and this  boosts
our confidence
in  our calculations. The predictions for $F_2$ for 
$\gamma^*$ - gold scattering for both approaches are given in Fig. \ref{F2Au},
again the results are close to one another.

\section{Discussion and Conclusions}   
\setcounter{equation}{0} 
 This paper demonstrates that assuming
 QCD  saturation of the parton densities 
at high energies,  one is able to describe  photoproduction processes, 
as well as DIS 
at small values of photon virtuality. We establish that the revised 
parametrization 
does not spoil the predictions made for
  DIS processes at large and moderate values of $Q^2$.  
This is   important as it allows us to use the DIS data to 
fix the phenomenological parameters of our approach.

 As in our treatment of the soft hadron interactions in 
Ref. \cite{BGLLM}, the present analysis  is based on
 two parameters and two major assumptions.
The parameters are:

(a) The effective radius of proton which is taken from the description of 
the DIS data for  $x<10^{-2}$; \\
(b) The parameter $Q_{0}^{2}$, (introduced in Eq. (\ref{newx})), 
  is adjusted by comparing with the high energy photoproduction 
data.

The two assumptions are:

(i) We use \eq{AN} as an ansatz for the $b_\perp$-dependence.  This 
Glauber-type
$b_\perp$ dependence has no solid theoretical foundation and should be 
considered to be an initial  guess.
It is therefore reassuring  to see
  that our
successful description of the photoproduction data is  in reasonable 
agreement  with this ansatz;
\\
(ii) We use hadron-parton duality to introduce the contribution of the 
secondary Regge trajectories.  The
 result that the  contribution of the secondary 
reggeons  is dual to the valence quark one, is unexpected and  
interesting.

Our main result, that we are successful in describing
 the photoproduction data for nuclear 
targets, is compelling
 and encourages us to suggest  further measurements of 
photon - nuclei interactions. We  stress that the Glauber approach 
for  nuclear targets does not contain any additional assumptions,  
thus  nuclear data can be used for clarifying the situation 
regarding our assumptions and fitting parameters.

 This paper and
 our previous work \cite{BGLLM}, provide
evidence  that an approach based on the QCD dipole picture and on QCD 
saturation is  successful in describing  processes which  
traditionally have been
associated with soft physics.  They also 
 reinforce the results obtained in
Ref. \cite{GLLM}   namely, that QCD saturation
 is a natural bridge between  soft physics and  conventional 
perturbative QCD.

\section{Acknowledgments} 
 
  We thank Nestor Armesto, Eran Naftali, and Anna Stasto for useful 
discussions.   
 E.L.  is indebted to the DESY Theory Division for their 
hospitality, while E.G. is grateful to the Department of Physics and 
Astronomy at UCI    for their cordial support.
M.L. thanks the Institute for Nuclear Theory at the University of 
Washington for its hospitality and the Department of Energy for partial
support during the completion of this work.
This research was supported in part by the 
GIF grant \# I-620-22.14/1999, and by the Israel Science Foundation, 
founded by the Israeli Academy of Science and Humanities.

\end{document}